\input{aipcheck}

\documentclass[
    ,final            
  ]
  {aipproc}

\layoutstyle{8x11single}

\begin{document}

\title{Gamma-Ray Burst Pulse Correlations as Redshift Indicators}


\author{Jon Hakkila}{
  address={Department of Physics and Astronomy, College of Charleston, Charleston, SC}
}

\author{P. Chris Fragile}{
  address={Department of Physics and Astronomy, College of Charleston, Charleston, SC}
}

\author{Timothy W. Giblin}{
  address={United Space Alliance/NASA JSC}
}

\begin{abstract}
Correlations among pulse properties in the prompt emission of long GRBs can potentially be used as cosmological distance indicators to estimate redshifts of GRBs to which these pulses belong. We demonstrate application of this technique to a sample of GRBs for which redshifts are not known. We also study the scatter of predicted redshifts of pulses found within individual bursts. We explore the characteristics of this scatter in hopes of identifying systematic corrections and/or pulse subsets that can be used to increase the technique's reliability.

\end{abstract}

\maketitle


\section{Introduction}

Pulses are the basic building blocks of gamma-ray burst (GRB) emission, and their correlative properties imply that GRB pulses are responsible for many luminosity-related GRB prompt emission characteristics. The lag vs.\ luminosity relation\cite{nor00}, the variability vs.\ luminosity relation\cite{rei01}, and the $E_{\rm pk}$ vs. $E_{\rm iso}$ relation \cite{ama02} all appear to result from {\em pulse} rather than {\em bulk} emission properties. Central to this argument is the recognition that each pulse has its own lag \cite{hak08a, hak08b}.  Since the pulse lag is directly related to pulse duration and since both anti-correlate with pulse peak luminosity, the bulk lag for a burst (obtained from the cross-correlation function \cite{ban97}) contains information about the pulse lags, while the burst's peak luminosity depends on some linear combination of pulse luminosities. The bulk and peak luminosity lag favor the highest intensity, shortest pulses, but smears out these measurements by sampling overlapping pulse properties rather than individual ones. The bulk lag vs.\ luminosity relation thus indicates a more fundamental pulse relation that has been distorted in a complex way via measurement of bulk properties.  Other correlations enabling luminosity calibration of GRBs are also distorted; these include the variability vs. luminosity relation \cite{rei01} and the $E_{\rm pk}$ vs. $E_{\rm iso}$ relation \cite{ama02}. Variability is related to the number of pulses in a burst, $E_{\rm pk}$ is a time-integrated value constructed from the merged spectra of many pulses, and the duration of a burst only reflects a pulse property when there is a single pulse; otherwise it reflects a mixture of pulse and interpulse durations. 

The centrality of GRB pulse luminosities suggests that the pulse properties can be used to determine GRB distances and redshifts. Using the pulse data collected by Hakkila and Cumbee (this conference) along with the calibrated pulse duration vs. \ pulse peak luminosity relation \cite{hak08a}, we study the hypothesis that pulse properties can be used to acquire GRB redshifts.

\section{Analysis}

We analyze 245 pulses in 106 Long GRBs measured by Hakkila and Cumbee (this conference), limiting our analysis to Long GRB pulses because the pulse peak lags and pulse durations of Long GRB pulses have been shown to anti-correlate with pulse peak luminosity, whereas correlations have not been demonstrated for Short GRB pulses. Pulse durations are preferred because they can be measured more accurately than pulse peak lags.

The relationship between pulse duration $w$ (interval between times where the intensity is $e^{-3}$ of the peak value on the 256 ms timescale in the 25 keV to 1 MeV energy range; $w_0$ refers to the rest-frame value) and isotropic pulse peak luminosity $L$ (measured in the same energy range and on the same timescale, in units of $10^{51}$ erg $s^{-1}$) has been found to obey a power law of the form 
\begin{equation}
L = A w_0^{-\beta}; 
\end{equation}
experimental values of $A$ and $\beta$ are $A = 1.53 \pm 0.02 $ and $\beta = -0.85 \pm 0.02$ \cite{hak08a}.

This correlative relationship can be combined with the definition of the luminosity distance $D_L$ to provide an independent method for determining the redshift $z$; this approach is similar to that used by Kocevski and Liang \cite{koc06} for estimating burst redshifts from the lag vs.\ luminosity relation. The definition of the luminosity distance is

\begin{equation}
D_L=\sqrt{\frac{L/d\omega}{p_{256}}}=(1+z)c/H_0\int_{0}^{z}\frac{dz}{\sqrt{\Omega_{m}(1+z)^3+\Omega_{\Lambda}}}
\end{equation}

where $d\omega$ is the beaming factor and standard cosmological parameters are assumed ($H_0=65$ km s$^{-1}$, $\Omega_m=0.3$, and $\Omega_\Lambda=0.7$). We assume for simplicity that beaming is isotropic, even though we know that this is not the case.

The equations can be solved iteratively to obtain the redshift for each pulse. When a burst has multiple pulses, the mean value of $z$ and a pulse spread error for the redshift measurement $\sigma_z$ can be obtained for the burst.

The redshifts of the measured bursts are shown in Figure 1a) as a function of the 256 ms peak flux$p_{256}$, with a mean pulse spread error of $\sigma_z \approx 0.7$ (for comparison, rough estimates of the errors in redshift obtained from Equation 1 are given by $\sigma_z \approx 0.2 z$). Few GRBs have $z>5$ (consistent with results of Ashcraft and Schaefer \cite{ash07}), implying that GRBs might have limited use in probing times as early as the Era of Recombination. We note that the brightest bursts are typically found at low-$z$, while the highest-$z$ bursts are faint (as measured by $p_{256}$). However, many faint bursts are also observed at low-$z$; these GRBs often have time history morphologies consistent with long-lag, single-pulsed bursts \cite{nor05, hak07}. Since pulse duration anti-correlates with pulse luminosity, it is not surprising that many low- and high-$z$ GRBs exhibit similar time history morphologies to each other (e.g., Figure 2 and Figure 3). 

Multi-pulse GRBs provide us with a method of checking the mean value of $\beta$: we assume that these pulses obey a power law function of the form $p_{256} = A w^{-\beta}$ within each burst. Multiple measurements of $\beta$ allow us to check if these values are consistent with the value of $\beta=0.8$ found for GRBs with known redshifts. Our sample contains 53 multiple-pulsed GRBs; from these we obtain the distribution shown in Figure 1b) and a mean power law index of $\beta = -0.8 \pm 0.4$ (error in the mean). Although this value is consistent with that obtained from pulses in GRBs with known redshifts, it is also consistent with no variation of duration with pulse luminosity. Removal of four large-residual outliers results in $\beta = -0.2 \pm 0.1$ (also error-in-the-mean). 

The small pulse spread error obtained for $\sigma_z$ obtained from GRBs without known redshifts using Equation 1, coupled with the lower value of $\beta$ obtained for the pulse sample without known redshifts relative to the pulse sample with known redshifts \cite{hak08a} suggests that {\em systematic} corrections are still needed. This is not surprising, because we started our analysis by purposefully ignoring two potentially important corrections: 1) K-corrections for redshifting the GRB spectra, and 2) non-isotropic pulse peak luminosities. Presumably, corrections of these types should result in a tighter fit of the data to the $L$ vs. $w$ relation. We have recently initiated studies of these effects, which have not to date been performed for {\em pulses}. We note in passing another difficult-to-quantify but potentially large source of error: our redshift calibration could be skewed by our inability to detect and fit faint, short duration, and/or overlapping pulses \cite{hak08a}.


\begin{figure}
  \includegraphics[height=.27\textheight]{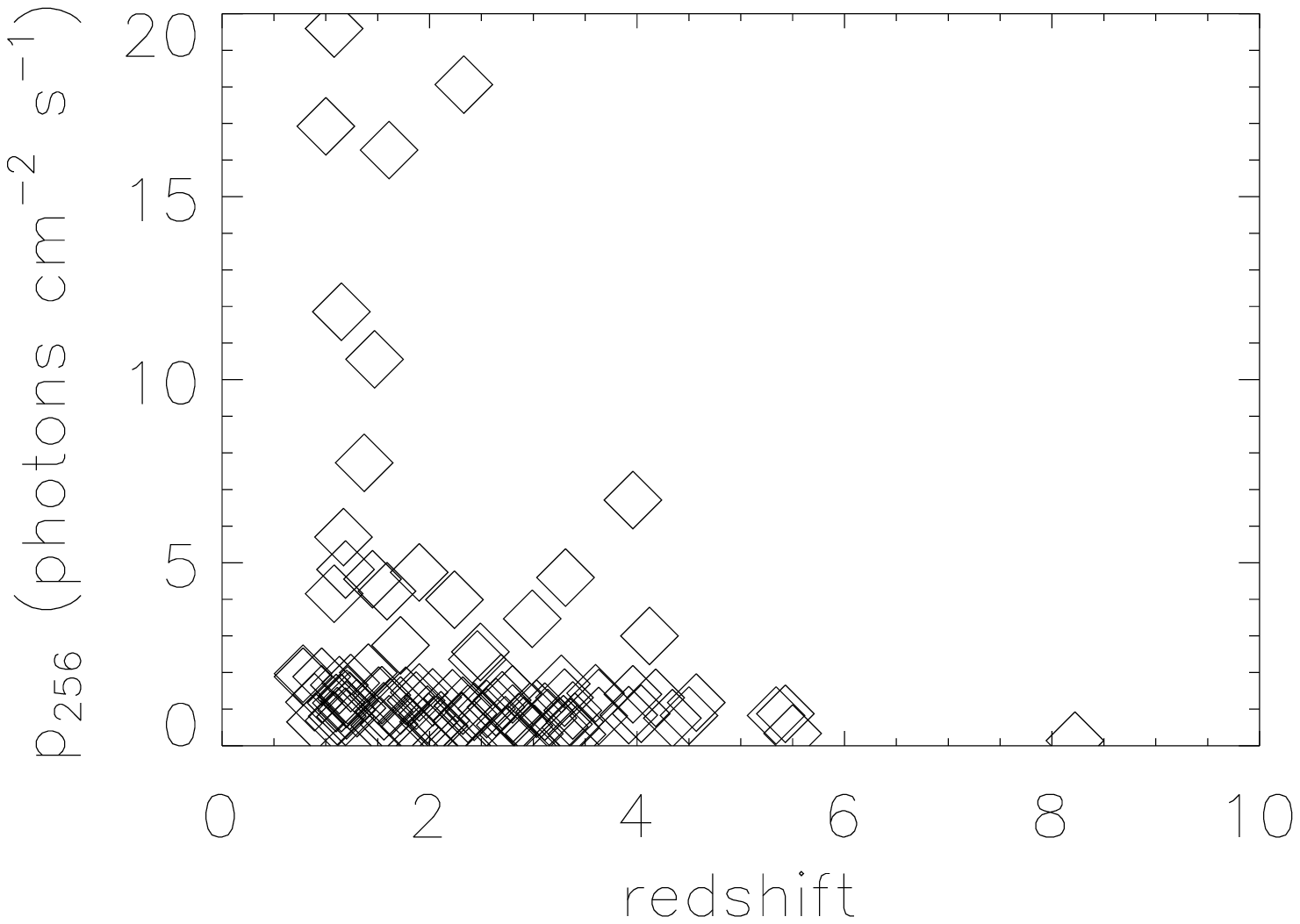}
  \includegraphics[height=.27\textheight]{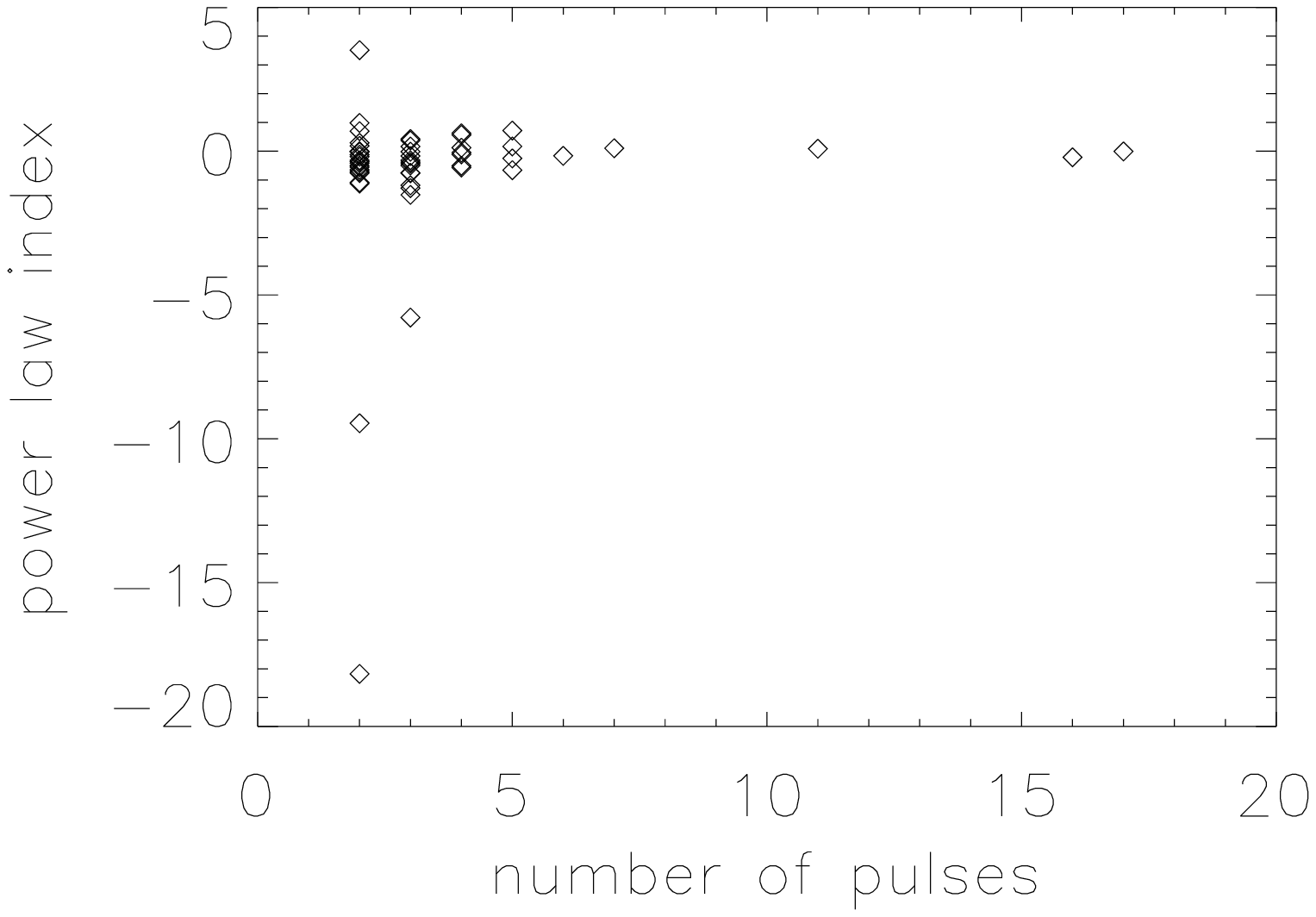}
  \caption{(a) 256 ms peak flux vs. redshift, as calculated from pulse durations. Few Long BATSE GRBs have $z>5$ -- note that the GRB pulse with $z \approx 8$ might be a misidentified Short GRB pulse. (b) Luminosity power law index $\beta$ obtained from GRBs with multiple pulses and fitting their peak fluxes and durations to a function of the form $p_{256} = A w^{-\beta}$.}
\end{figure}

\begin{figure}
  \includegraphics[height=.22\textheight]{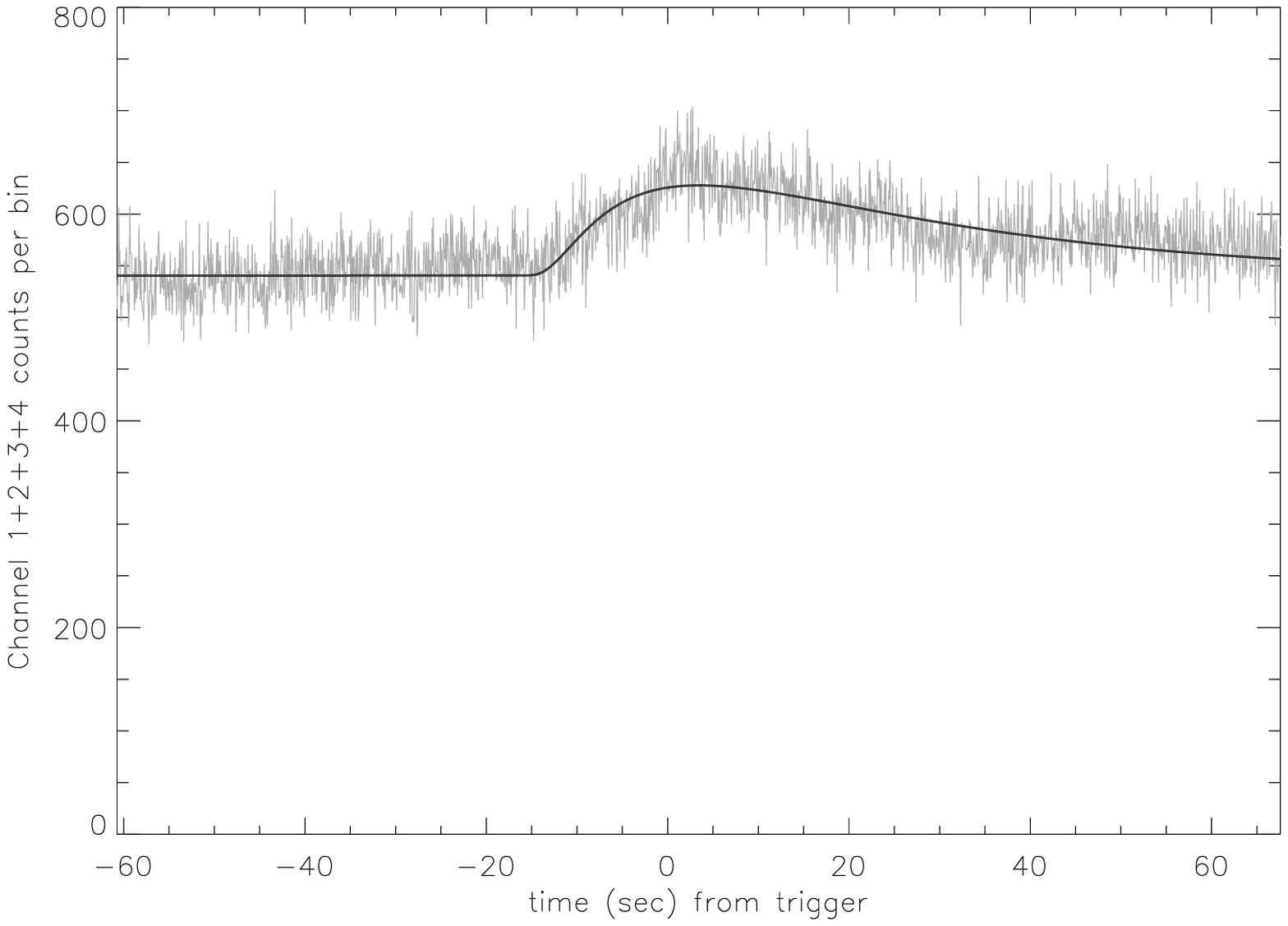}
  \includegraphics[height=.22\textheight]{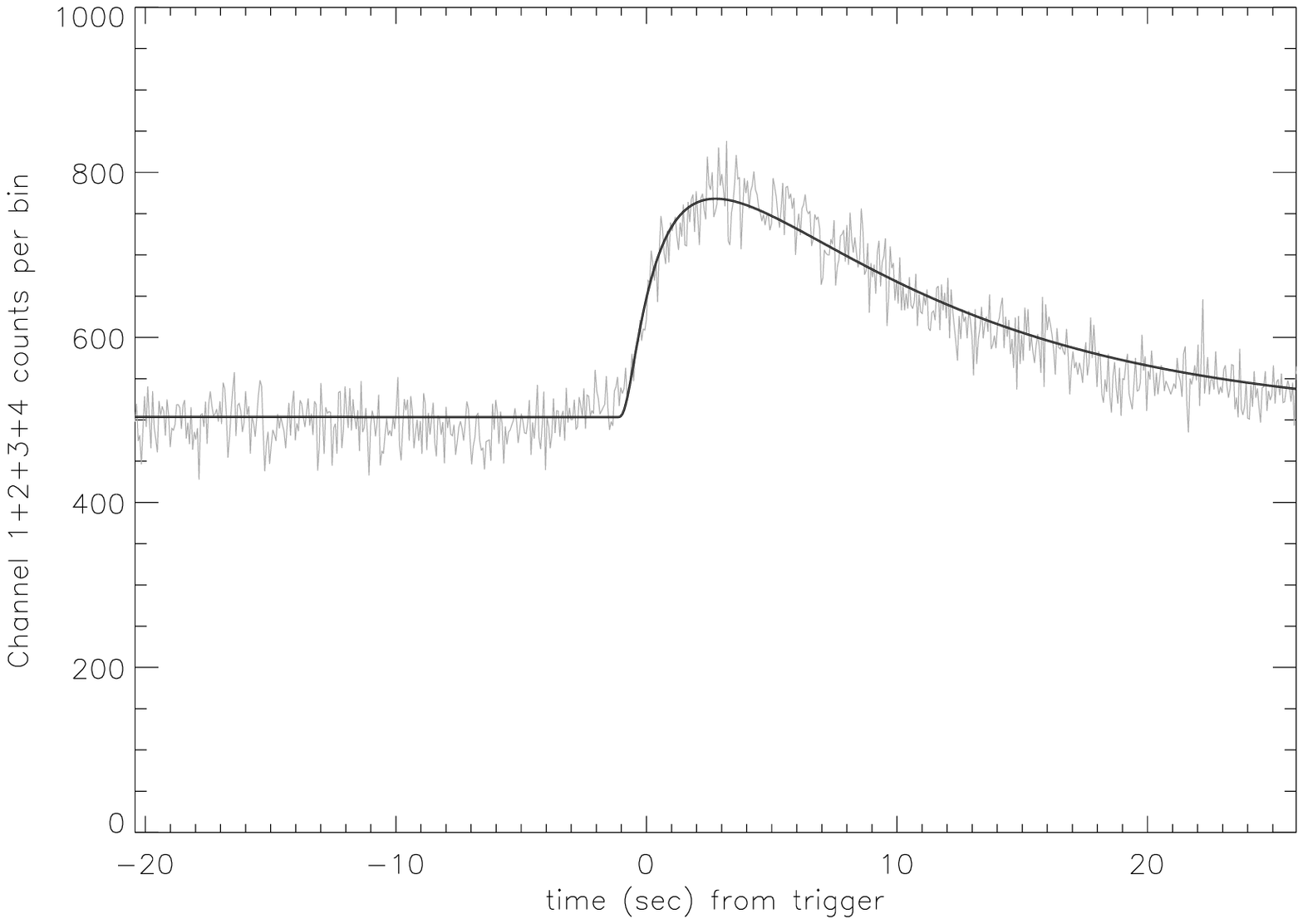}
  \caption{Examples of low-$z$ GRBs (a) BATSE trigger 111 ($z=0.9$). (b) BATSE trigger 332 ($z=0.9$).}
\end{figure}

\begin{figure}
  \includegraphics[height=.22\textheight]{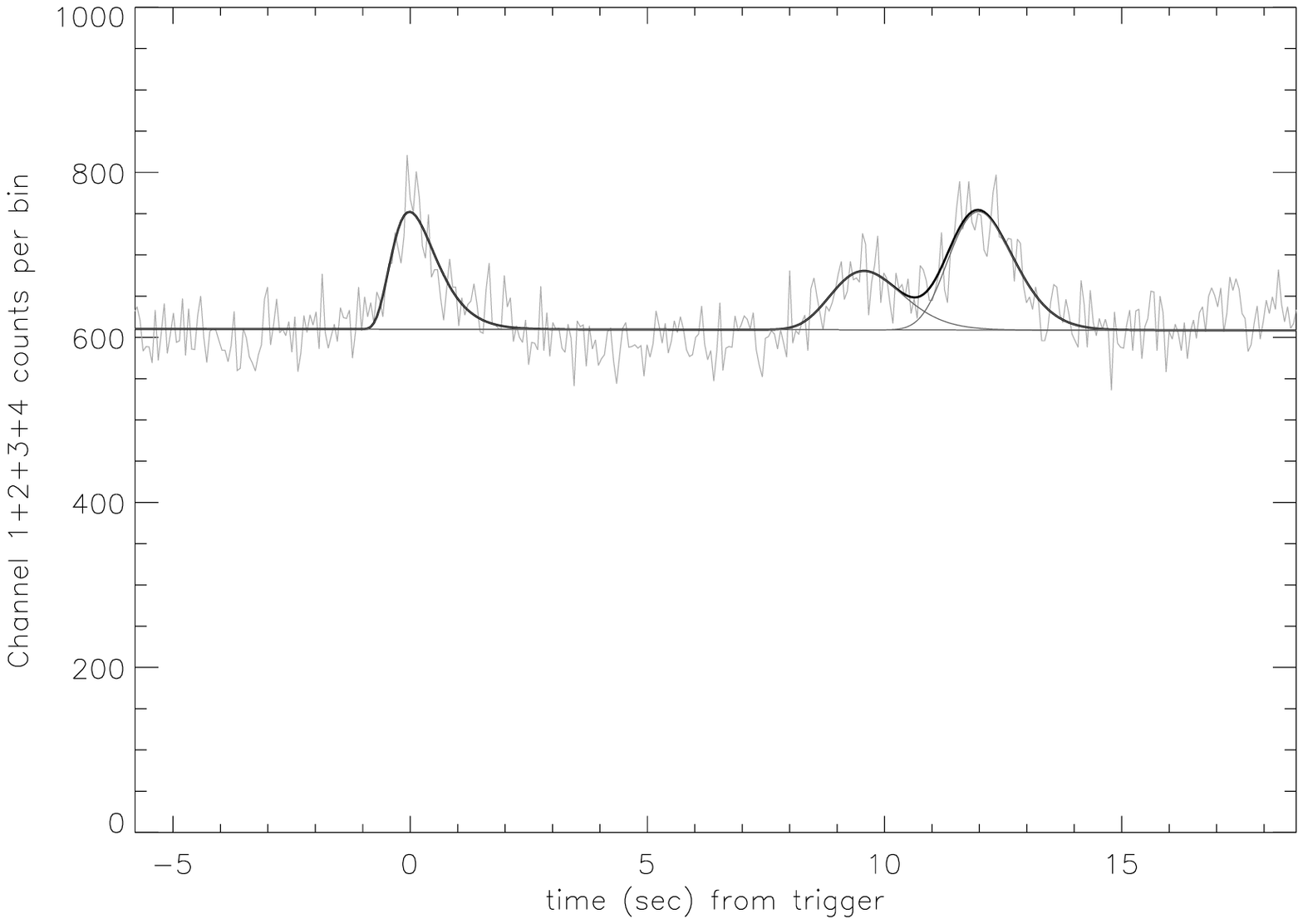}
  \includegraphics[height=.22\textheight]{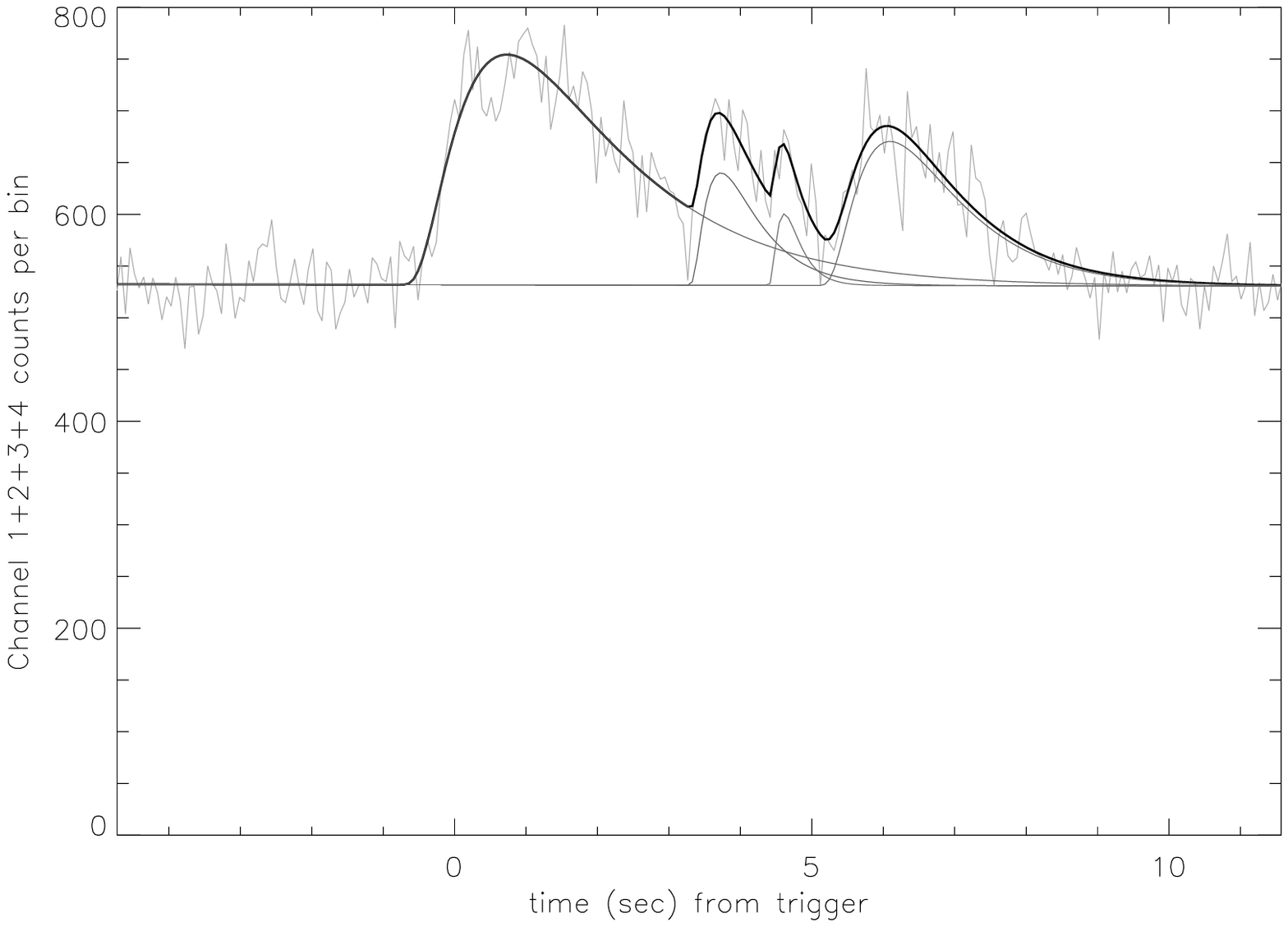}
  \caption{Examples of high-$z$ GRBs (a) BATSE trigger 214 ($z=4.3$). (b) BATSE trigger 803 ($z=4.6$).}
\end{figure}

\section{Conclusions}

\begin{itemize}
\item Pulse durations (and, less accurately, pulse lags) can be used to obtain GRB redshifts. Variations in individual pulse luminosities can be minimized by obtaining an average over pulses within a burst. 
\item BATSE GRBs typically have $z<5$, with few high-$z$ GRBs observed.
\item The predictive capability of the technique described here can be improved by correcting for cosmological shifting of pulse spectral characteristics and by knowing pulse beaming factors.
\end{itemize}





\bibliographystyle{aipprocl} 

\bibliography{pulse_props}

\IfFileExists{\jobname.bbl}{}
 {\typeout{}
  \typeout{******************************************}
  \typeout{** Please run "bibtex \jobname" to optain}
  \typeout{** the bibliography and then re-run LaTeX}
  \typeout{** twice to fix the references!}
  \typeout{******************************************}
  \typeout{}
 }

\end{document}